\documentclass{article}
\usepackage[utf8]{inputenc}
\usepackage{graphicx,epsfig}
\usepackage{amsmath, hyperref}
\usepackage{amsfonts}
\usepackage{fancyhdr}
\usepackage{subcaption}
\usepackage{authblk}
\usepackage{relsize}
\usepackage[margin=2.7cm]{geometry}
\newcommand{\subtitlerelsize}{3} 
\newcommand{\subtitlelinesep}{0.2em} 

\title{Protein Unfolding and Aggregation near a Hydrophobic Interface\\[\subtitlelinesep]%
    \smaller[\subtitlerelsize]{}Print submitted to MDPI Polymers, January 2021}

\author[1]{David March}
\author[2]{Valentino Bianco}
\author[1]{Giancarlo Franzese}
\affil[1]{Secci\'o de F\'isica Estad\'istica i Interdisciplin\`aria---Departament de F\'isica de la Mat\`eria Condensada, \mbox{Facultat de F\'isica,} Universitat de Barcelona, Mart\'i i Franqu\`es 1, 08028 Barcelona, Spain}
\affil[2]{Chemical Physics Department, Faculty of Chemistry, Universidad Complutense de Madrid, \mbox{Plaza de las Ciencias}, Ciudad Universitaria, 28040 Madrid, Spain}
\date{}
\begin{document}
\maketitle
\abstract{The behavior of proteins near interfaces is relevant for biological and medical purposes. Previous results in bulk show that, when the protein concentration increases, the proteins unfold and, at higher concentrations, aggregate. Here, we study how the presence of a hydrophobic surface affects this course of events.  To this goal, we use a coarse-grained model of proteins and study by simulations their folding and aggregation near an ideal hydrophobic surface in an aqueous environment by changing parameters such as temperature and hydrophobic strength, related, \mbox{e.g., to ions} concentration.  We show that the hydrophobic 
surface, as well as the other parameters, affect both the protein unfolding and aggregation.
We discuss the interpretation of these results and define future lines for further analysis, with their 
 possible implications in neurodegenerative diseases.}

\section{Introduction}
Proteins cover a range of fundamental actions in a living organism, including enzymatic and hormone functions, transport of biomolecules within the cellular environment, energy sourcing, tissues build and repair~\cite{Finkelstein:2016aa}. Typically a protein can perform these functions only when it is in its native and folded~conformation.  

Protein folding is a self-organized process occurring spontaneously in the aqueous solution, at~least for small proteins, and~it is dictated mostly by the protein sequence.  
After~the synthesis at the ribosome, the~polypeptide chain finds itself in a highly crowded cellular environment. Here, despite many non-specific interactions, the~chain is capable of selecting a subset of amino acid contacts that funnel the free energy landscape toward a unique native/folded~state.

However, the~accumulation of partially folded conformations or the competition with other unfolded proteins could hinder the folding process, resulting in the formation of macromolecular aggregates~\cite{Kiefhaber:1991aa}. 
Proteins can aggregate after they fold in the native state through chemical bonding or self-assembling---or via unfolded intermediate conformations. In~particular, non-native protein-aggregates are commonly formed through a multi-step process and are made of native-like--partially-folded intermediate structures~\cite{Eliezer:1993aa, Fink:1998aa, Roberts:2007aa, Neudecker362}.

Proteins have a propensity to aggregate related to a series of factors, e.g.,~the flexibility of the protein structure~\cite{De-Simone:2012ve} or the sub-cellular volume where the protein resides~\cite{B913099N}.
They~evolved toward a low aggregation-propensity, within~a range of protein expression required for their bioactivity. 
However, they have no margin to respond to external factors that increase or decrease their expression or solubility~\cite{Schroder:2002aa, Tartaglia:2007aa, B913099N}.
As a consequence, inappropriate protein aggregation represents a crucial issue in biology and medicine. It is associated with a growing number of diseases, such as Alzheimer's and Parkinson's disease~\cite{Ross:2005aa, Chiti:2006aa, Aguzzi:2010aa, Knowles:2014aa}, and~with the degradation of the pharmaceutical product quality and performance~\cite{Roberts:2014aa}.

Among different strategies to tackle the related diseases, many hopes have been placed in using functionalized nanoparticles for inhibiting protein and peptide aggregation~\cite{Zaman:2014aa, Howes1247390}.  However, once in the bloodstream, the~nanoparticles form the protein corona~\cite{Vilanova2016, C9CP06371D}. This corona can alter the biological effect of the nanoparticle and can induce unexpected reactions~\cite{Dobrovolskaia:2007aa, Qing:2014aa}. 

Many aspects of the nanoparticle interface, such as the shape, the~size, or~the surface chemistry, can affect the aggregation of proteins~\cite{Hook:1998aa, Henry:2003aa, Lacerda:2010aa, Canaveras:2012aa, C3SM27710K, Wei:2017aa}. Nonetheless, the~capability of proteins to keep their native conformation upon aggregation or adsorption onto inorganic interfaces is still poorly~understood. 

Computational approaches are gaining ground as fundamental tools to investigate these phenomena and the interplay between folding and aggregation in homogeneous and heterogeneous solutions of proteins. In~particular, coarse-grain models and multiscale methods allow us to deal with such complex systems~\cite{Cellmer:2007aa, Nasica-Labouze:2015aa}. 

For example, a~recent study showed that the concentration increase of individual protein species can unfold their native state without inducing their aggregation~\cite{Bianco:2020aa}. \mbox{Furthermore,~each} component in a protein mixture can keep its folded state at densities that are larger than those at which they would precipitate if they were in a single-specie
 solution~\cite{Bianco-Navarro2019}. 
 
These works study a lattice model of proteins embedded in an explicit coarse-grain water-model. The~models account for the protein features~\cite{Franzese2011, Bianco2012a, FranzeseBiancoFoodBio2013, Bianco:2015aa, Bianco:2017aa, BiancoPRX2017} and the water thermodynamics~\cite{FranzeseSCKMCS2008, Stokely2010, MSPBSF2011, delosSantos2011, Bianco2014, Bianco:2019aa}, representing a promising approach to study the behavior of protein solutions at an inorganic interface. 

This approach has been exploited to better understand 
the mechanisms of cold and pressure denaturation of proteins and the effect of 
 water-mediated interactions. 
\mbox{Taking~into} account how water at the protein interface changes its hydrogen bond properties and its density fluctuations,
the model can 
 predict protein stability regions with elliptic shapes in the temperature-pressure plane, consistent with other theories
and experiments, \mbox{identifying~the} different mechanisms with which water participates to denaturation by changing temperature or pressure~\cite{Bianco:2015aa}.
 
Furthermore, this model can be used to design proteins at extreme conditions of temperature and pressure.
It has clarified that the limits of stability in temperature and pressure, and~the selection mechanisms at extreme conditions, relate to the temperature and pressure dependence of the properties of the surrounding water~\cite{BiancoPRX2017}. As~a consequence, the~hydropathy profile of the proteins results from a selection process influenced by water, with~superstable proteins at high temperatures characterized by nonextreme segregation between the hydrophilic surface and the hydrophobic core, while less-stable proteins have larger segregation or very low segregation~\cite{BiancoPRX2017}.

Here, following the approach in Ref.~\cite{Bianco:2020aa, Bianco-Navarro2019}, we present a computational study on the protein folding/unfolding and aggregation near a hydrophobic interface, representative of a portion of a nanomaterial. By~performing Monte Carlo simulations, we describe the formation of aggregates against folded conformations at different temperatures, both in the bulk water and at the hydrophobic interface.  We discuss the dependence of our finding on the water--water interaction in the protein hydration shell, linking the observed phenomena to the hydrophobic effect. 
Our results could shed light on the biological mechanisms underlying the formation of protein aggregates at the~nanoscale.

\section{Model}
\subsection{Franzese--Stanley Water~Model}
  We adopt a coarse-grain representation of the water molecules, partitioning a volume $V$ into a fixed number $N$ of cells, each one with volume $ v\equiv V/N \geq v_0$, with $v_0$ being the water excluded volume. For~the sake of simplicity, we will consider here the  case of the projection into two dimensions (2D) of a water monolayer with height $h\simeq 0.5$  nm. Although~a confined monolayer of water can have properties quite different from bulk water~\cite{Gopinadhan145, calero2020}, here the dimensionality only affects the number of neighbors of each water molecule but does not change its coordination number (the number of hydrogen bonds formed by each water molecule). Indeed, regardless of whether the model is in 2D or 3D, each water molecule can form up to four hydrogen bonds. Our preliminary data show that this is sufficient to find no qualitative differences near ambient conditions between our water model in 2D and 3D~\cite{Coronas2016}.
  
  We fix $T$ and $P$ of the system, leaving $r\equiv \sqrt{v/h}$ free to change, with~$r \geq r_0$ the average distance between first neighbor water molecules. The~model is able to describe all the fluid phases of water~\cite{Bianco2014}. Here we focus only on its liquid phase. The~Hamiltonian describing the interaction of the bulk water is
\begin{equation}
      \mathcal{H}\equiv \sum_{ij}\mathcal{U}(r_{ij})-J N_{HB}^{(b)}-J_{\sigma}N_{\rm coop}^{(b)}.
\label{eq:H}
  \end{equation}
  
  The first term, summed over all the molecules $i$ and $j$ at oxygen--oxygen distance $r_{ij}$, accounts for the Van der Waals attraction and the repulsive forces due to Pauli’s exclusion principle, and~is expressed as a double-truncated Lennard--Jones potential,
    \begin{equation*}
      \mathcal{U}(r)\equiv 4 \epsilon \left[ \left( \frac{r_0}{r} \right)^{12} - \left( \frac{r_0}{r} \right)^{6} \right], \quad \; \text{if } r_0<r<6 r_0,
  \end{equation*}
  
  $\mathcal{U}\equiv \infty$ for $ r \leq r_0 $ and $\mathcal{U}\equiv 0$ for $r\geq 6 r_0$, where we use  $\epsilon$ as our energy~scale. 
  
 The second term of the Hamiltonian represents the directional (covalent) contribution to the formation of water--water hydrogen bonds (HBs) with characteristic energy $J$. \mbox{Assuming that} each molecule $i$ can form up to four HBs, the~number of possible molecular conformations is discretized by the introduction of four bonding variables $\sigma_{ij}=1, \dots, q$, one~for each neighbor molecule $j$. Following a standard definition~\cite{Luzar-Chandler96}, two conditions must hold for the formation of a~HB. 
  
First, the~molecules must be separated no further than $r_{\rm max}$.  In~a monolayer the condition  $r< r_{\rm max}$ corresponds to 
$v/v_0<0.5$ for  $v_0=r_0^2h$, with~$r_0\simeq 2.9$~\AA\ van der Waals diameter of a water molecule, and~$r_{\rm max}\simeq 4$~\AA. We associate to each water molecule $i$ with a proper volume $v$ an index $n_i=1$ if $v/v_0<0.5$, and~$n_i=0$ otherwise. Hence, for~the neighbor molecules $i$ and $j$,  the first necessary condition to form a HB is that $n_in_j=1$.

Second, the~angle $\widehat{OOH}$
 between two neighbor molecules must be less than $\pm$30º. Therefore, only 1/6 of all the possible orientations [0º, 360º] are associated with a HB. Thus, we fix $q=6$, and~the second condition to form 
 a HB is that $\sigma_{ij}=\sigma_{ji}$,  correctly accounting for the entropy loss associated with a HB formation. 
 Therefore,
 the total number of bulk HBs is $N_{HB}^{(b)}\equiv \sum_{\langle ij\rangle} n_in_j\delta_{\sigma_{ij},\sigma_{ji}}$, where $\delta_{a,b}=1$ if $a=b$, 0 otherwise, and~ the sum is over nearest neighbor~molecules.
 
The third term of Eq.~(\ref{eq:H}) corresponds to the cooperative interaction of the HBs, emerging from quantum many-body interactions, which leads to an ordered, low-density tetrahedral configuration in bulk. This phenomenon is modeled as an effective interaction between each of the six different pairs of the four variables $\sigma_{ij}$ of a molecule $i$, coupled by an energy $J_{\sigma}$. $N_{\rm coop}^{(b)}\equiv \sum_i n_i \sum_{kl} \delta_{\sigma_{ik},\sigma_{il}}$ is the sum over the pair of bonding indices that cooperatively acquire the same value in each molecule $i$. By~taking $J_{\sigma} \ll J$, we guarantee that the term plays a role only when the HBs are~formed. 
  
  Finally, the~total volume, and~hence the density field,  depends on the HB formation, as~$V^{(b)}\equiv Nv+N_{HB}^{(b)} v_{HB}^{(b)}$, where $v_{HB}^{(b)}$ is a fraction of $v_0$. This relation accounts, on~average, for~the local decrease of density due to the tetrahedral HB network. The~values of the model's parameters  are given at the end of the next section. Further details about the water model can be found in Ref.~\cite{Coronas:2021aa}.
  
\subsection{Protein and Interface~Model}
Following the coarse-grain representation for the water molecules, we adopt a coarse-grained lattice representation for the proteins, depicted as self-avoiding heteropolymers composed of 36 amino acids.  For~ simplicity, each residue can occupy only one of the cells of the~system.

The amino acids interact through the nearest neighbor potential given by the Miyazawa--Jerningan interaction matrix~\cite{Miyazawa:1985aa}.
To account for the lower surface-volume ratio in 2D, we~scale the matrix by a factor of 2, increasing the effective amino acids interactions~\cite{Bianco:2020aa}. 
 

Depending if two water molecules, forming a HB, are near two hydrophobic ($\Phi$)  amino acids, two hydrophilic ($\zeta$)  amino acids, or~one of each kind (mixed, $\chi$), the~hydration-water Hamiltonian is
\begin{equation} 
       \mathcal{H}_{w,w}^{(h)} \equiv - \left[ J^{\Phi} N_{HB}^{\Phi} + J^{\zeta} N_{HB}^{\zeta} + J^{\chi} N_{HB}^{\chi} \right] 
       - \left[ J_{\sigma}^{\Phi} N_{\rm coop}^{\Phi} + J_{\sigma}^{\zeta} N_{\rm coop}^{\zeta} + J_{\sigma}^{\chi} N_{\rm coop}^{\chi} \right],
   \end{equation}
where $N_\alpha^{\Phi}$, $N_\alpha^{\zeta}$, $N_\alpha^{\chi}$ ($\alpha=HB, {\rm coop}$) represent the number of directional and cooperative bonds formed at a hydrophobic, hydrophilic or mixed interface, respectively. 

Experiments and simulations  show that the water--water HBs near a hydrophobic interface are (i) stronger than bulk HBs, and~(ii) increase the local water density upon pressurization~\cite{Sarupria:2009ly}. 
To account for these effects, the~model assumes  that $J^{\Phi}>J$, $J^{\Phi}_{\sigma}>J_{\sigma}$, and~ that the volume associated with a HB at the $\Phi$ interface decreases upon a pressure $P$ increment, i.e.,
\begin{equation} 
v^{\Phi}_{HB} / v^{\Phi}_{HB,0}\equiv 1-k_1 P,
   \end{equation}
where $v_{HB,0}^{\Phi}$ is the volume increase for $P=0$, and~$k_1$ is a factor accounting for the compressibility of the hydrophobic hydration shell. 
Thus, HBs in a hydrophobic hydration shell generate an extra contribution $V^{\Phi}\equiv N_{HB}^{\Phi} v_{HB}^{\Phi}$  to the total~volume.

We adopt the simplified version of the model in which the HBs and the water density near a hydrophilic interface are as in bulk. The~parameters for the HBs in the mixed, $\chi$, case are an average between the  $\Phi$ and the  $\zeta$ case.
Hence, the~model  sets $J^{\zeta}=J$, $J_{\sigma}^{\zeta}=J_{\sigma}$,  $J^{\chi}=(J^{\Phi}+J^{\zeta})/2$, $J_{\sigma}^{\chi}=(J_{\sigma}^{\Phi}+J_{\sigma}^{\zeta})/2$, and~lastly
   $v_{HB}^{\zeta}=v_{HB}^{(b)}$, where $v_{HB}^{(b)}$ is the bulk~HB-volume parameter. 
   
   Finally, the~model assumes that the protein--water isotropic interaction energy is different depending on the residue nature. In~particular, it is  $- \varepsilon^{\Phi}$ and $- \varepsilon^{\zeta}$ in the hydrophobic and the hydrophilic hydration shell, respectively.

\subsection{The Hydrophobic~Surface}

We model the hydrophobic interface as a flat surface with excluded-volume interaction with both water and proteins. We fix it in space, separating our systems into two parts. 
Because we consider periodic boundary conditions (PBC), our system corresponds to an infinite volume confined between two parallel hydrophobic surfaces at a distance equal to the size $L$ of the system. The~water--water HBs near the hydrophobic interface are as in the  $\Phi$ case described~above.

\begin{table}[htp]
\caption{The three sets of parameters--here called Scales--considered in this work 
for the coarse-grained proteins, hydrated by the Franzese-Stanley water model. 
The three sets differ only for the values of $\varepsilon^{\Phi}$, $J^{\Phi}$, and $J_{\sigma}^{\Phi}$, associated with the water hydrating hydrophobic interfaces/residues.
Symbols are defined in the text.}
\begin{center}
\begin{tabular}{||c | c | c | c | c | c | c | c | c | c ||}
\hline
Scale & $J/8 \varepsilon$ & $J_{\sigma}/8 \varepsilon$ & $v_{HB}^{(b)}/v_0$ & 
$\varepsilon^{\Phi}/8 \varepsilon$ & $J^{\Phi}/8 \varepsilon$ & $J_{\sigma}^{\Phi}/8 \varepsilon$ &  $v_{HB,0}^{\Phi}/v_0$ & $k_1\varepsilon/v_0$ &
$\varepsilon^{\zeta}/8 \varepsilon$ \\ 
\hline
\hline
0 & $0.3$ & $0.05$ & $0.5$ & 
$0.48$ & $1.2$ & $0.2$ & $2$ & $4$ &
$0$ \\
\hline
1 & $0.3$ & $0.05$ & $0.5$ & 
$0.24$ &$0.6$ & $0.1$ &  $2$ & $4$ &
$0$ \\
\hline
2 & $0.3$ & $0.05$ & $0.5$ & 
$0.1$ &$0.35$ & $0.05$ & $2$ & $4$ &
$0$ \\
\hline
\end{tabular}
\end{center}
\label{Tab1}
\end{table}%

\subsection{The Model's~Parameters}
Following Ref.~\cite{Bianco:2020aa}, we choose model's parameters that 
balance the water--water, water--residue, and~residue--residue interactions, ensuring the protein stability in the liquid phase, including ambient conditions.
Furthermore, by~enhancing the interfacial interactions, they  also account for the protein and interface surface loss by taking a 2D representation instead of 3D.   
As described in the following, we use three different sets of parameters to understand how our results depend on them. The~three sets, here called Scale 0, 1, and~2, are indicated in Table~\ref{Tab1}. 
For the sake of comparison, Scale 0 is the same set of parameters adopted in Ref.~\cite{Bianco:2020aa}.

\section{Method}

\subsection{The~Protein}
\begin{figure}
     \centering
     \includegraphics[scale=0.34]{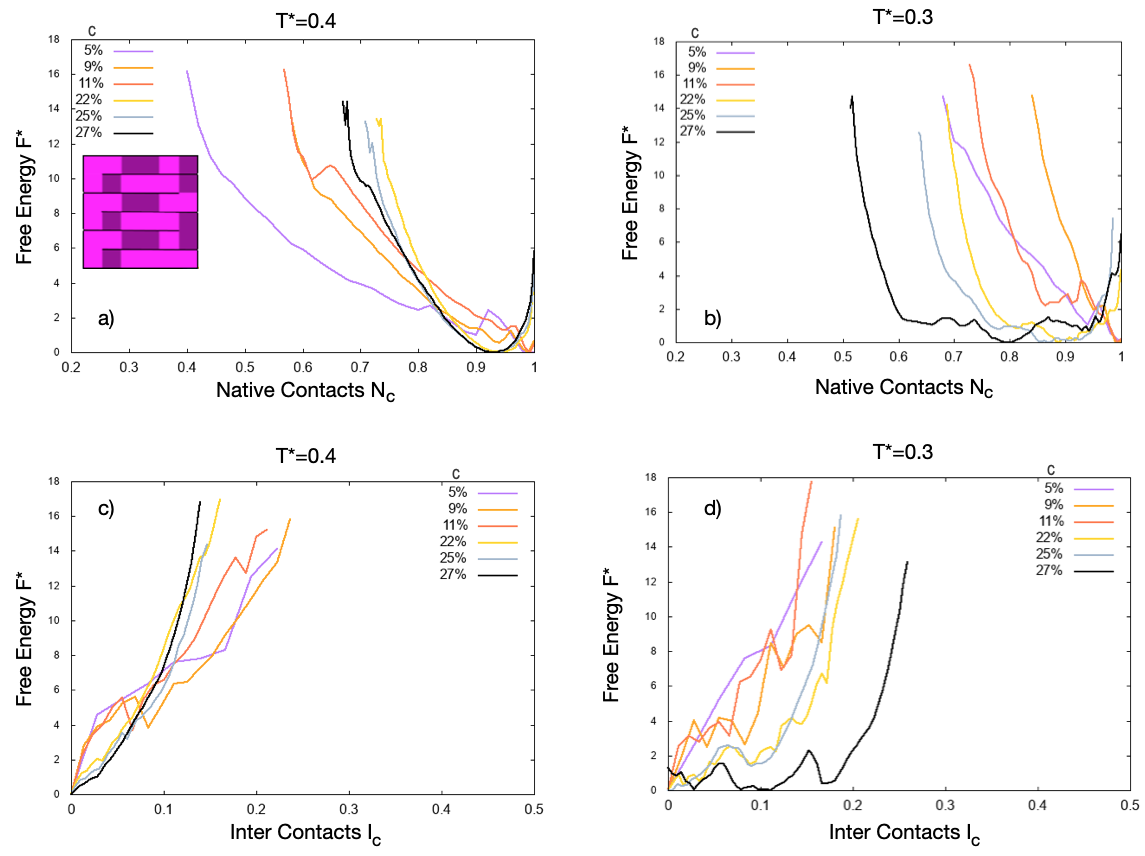}    
           \caption{The  {\it snake}-proteins free energy changes with concentration $c$ and temperature $T^*$. 
           The concentrations go from $c=5\%$ (violet) to $27\%$ (black), as~indicated in the legend.
           (\textbf{a}) Inset: Native structure of the {\it snake} protein; the dark/light cells are hydrophobic/hydrophilic amino acids. Main panels: 
           $F^*(N_c)$  at warm temperature ($T^*=0.4$), and~
           (\textbf{b}) at ambient temperature ($T^*=0.3$), has a minimum that moves from
           $N_c=1$ at low $c$ to $N_c<1$ at high $c$, with~a change between the folded and the unfolded state at $11\%<c<22\%$, corresponding to 
           $5\leq N_p\leq 10$ proteins in our system.  The~change is greater at ambient conditions. 
            (\textbf{c}) For warm temperature and the same concentrations, $F^*(I_c)$ has a minimum always at $I_c=0$, showing that the proteins do not aggregate at this temperature, regardless if they are folded or unfolded. 
          (\textbf{d})~The~result is different at ambient conditions. The~free energy develops several minima at the larger concentration, $c=27\%$ (12~proteins), showing that the unfolded proteins aggregate at high concentration. 
 In all the panels, the~curves at lower concentrations are noisier than those at higher $c$ because the corresponding averages are over smaller numbers of proteins. Hence, the~fluctuations along the curves are an indication of the error bar on the estimates.
           In general, a~detailed study of the free-energy landscape to estimate the possible occurrence of free-energy barriers would imply much larger statistics, which is out of the scope of the present work.} 
       \label{fig:NcIc}
   \end{figure}

We study proteins with a {\it snake}-like native state (Figure~\ref{fig:NcIc}, a~inset). For comparison, we choose the $A_0$ protein introduced in Ref~\cite{Bianco:2020aa}, which is in its native state at ambient conditions. Each protein has 36 residues and a
hydrated interface of 20 amino acids of which 
7 ($35\%$) are hydrophobic and 
13 ($65\%$) hydrophilic.
It has one side fully hydrophilic and
no side completely~hydrophobic. 

\subsection{The Monte Carlo~Simulation}

We perform Monte Carlo (MC) simulations of  {\it snake} proteins 
embedded into  a square lattice, with~size $L=40$ and PBC.
We consider protein concentrations in the range 
\mbox{$c=[4.5\%, 27\%]$} in volume,
from $N_p=2$ to 12 proteins.
We simulate the system at ambient conditions, as~in Ref.~\cite{Bianco:2020aa}, and~at a warmer temperature.
In internal units of the coarse-grained water model, these thermodynamic conditions correspond qualitatively to 
set $ k_B T / \varepsilon=0.3$ for ambient conditions, and~$ k_B T / \varepsilon=0.4$ for warmer water, both at $P=0$.

Following Ref.~\cite{Bianco:2020aa}, each MC step is defined by the following~sub-steps: 
\begin{enumerate} 
       \item We choose randomly a global protein-move among shift, rotation, crankshaft, or~pivot~\cite{Frenkel2002b}. 
Then, we pick at random one of the proteins, and~we attempt the selected global move. 
We repeat the random selection for $N_p$ times, updating on average all the proteins.
       \item We choose a random number $m$ between 1 and $4L^2$. 
For $m$ times,  we select one of the $L^2$  cells. 
If it includes an amino acid, we attempt a corner flip, i.e.,~the local protein-move~\cite{Frenkel2002b}.
If it includes a water molecule, we select one of its four $\sigma$-variables and attempt to change its state, hence, breaking or forming a HB.
	\item We attempt a global change of the system volume.
   \end{enumerate}
   
We accept or reject each step following the MC detailed-balance rules.   
This algorithm guarantees that, for~each possible global change in the protein configurations, there is a random number of local moves for the proteins or the water. This choice allows the system to re-equilibrate during the~process.  

\subsection{The~Observables}
 
To study the proteins folding/unfolding and aggregation, at~each MC step,  we calculate 
the number $N_c$ of native contacts of each protein, i.e.,~contacts in common with the native structure,
normalized by its maximum value ($25N_p$).
Furthermore, we compute 
the number $I_c$ of inter-contacts between different proteins, or~between  proteins and the interface,
normalized by its maximum ($36N_p$).
Finally, we calculate 
the number $M_c$ of contacts of the proteins with the interface, normalized by ($2L$).

For each $c$ and $T$, first, we equilibrate the system. We start from a high-$T$ configuration for water, where we distribute the proteins in a homogenous way in their extended configuration.
We consider that the system is equilibrated when all the observables $N_c$, $I_c$, and~$M_c$ at each MC step fluctuate around an average value without displaying any drift at least for $10^6$  MC steps. 
We observe that,  for~Scale 0 parameters  (Table~\ref{Tab1}), $1\times 10^6 \div 10\times 10^6$  MC steps provide enough time to reach equilibrium, depending on $c$ and $T$. The~equilibration time is longer for larger $c$ and smaller $T$.
For Scale 1 and 2, which we studied only at concentration 11\%, the~system slows down, requiring   $12\times 10^6$  MC steps of~equilibration.

Once at equilibrium, we calculate during  $10^7$  MC steps the probability of occurrence, $P(O)$,  
for each observable $O$. 
Finally,
we compute the free-energy as function of each observables,  $F(O)\equiv -k_B T \ln{P(O)}$.
In the following, we use dimensionless temperature 
$T^{*} \equiv k_B T / \varepsilon$ and  free-energy
$F^{*} \equiv F / \varepsilon$.

\section{Results}

\subsection{Scale~0}

We first simulate the system at different concentrations and warm temperature. 
The~free energy as a function of the normalized number of native contacts, $F^*(N_c)$, shows~a minimum near $N_c=1$ at low concentration, $c\leq 11\%$ (Figure~\ref{fig:NcIc}a).
In this regime, all the proteins fold in their native conformation. 

At higher concentrations, $c>11\%$,  the~minimum moves toward $N_c\simeq 0.94$. Hence,  the~proteins are on average in slightly-unfolded states, with~$\sim 94\%$ of their native contacts.  

The unfolding at $c>11\%$ is more evident at ambient temperature (Figure~\ref{fig:NcIc}b). As~the concentration increases, the~$F^*$ minimum moves toward lower values of $N_c$ and becomes broader. Both changes indicate a larger propensity of the proteins to unfold toward configurations with less than 80\% of their  native contacts for increasing $c$. 

We calculate, under~the same conditions, the~tendency of the proteins to aggregate or adsorb onto the hydrophobic surfaces. 
At the higher temperature, the~proteins, on~average, do not aggregate or adsorb at the concentrations considered here. This result is clearly shown by the free energy  as a function of the normalized number of inter contacts, $F^*(I_c)$, 
with minima at $I_c^*=0$ at any $c$ (Figure~\ref{fig:NcIc}c).

However, the~situation changes at ambient conditions (Figure~\ref{fig:NcIc}d). By~increasing the concentration, for~ $c>25\%$, above~the unfolding  threshold $c>11\%$,
 they aggregate or adsorb onto the hydrophobic walls, with~a shallow minimum in $F^*(I_c)$ around $I_c\simeq 0.1$ for $c=27\%$. This minimum shows that,
 when the native contacts are less than 80\%, more than 10\% of the total volume of the proteins is  aggregated or adsorbed at a high concentration.   Further calculations clarify that this minimum is only due to protein--protein aggregation, with~no contribution from~surface-adsorbed proteins.

\begin{figure}[h]
    \centering
     \includegraphics[scale=0.35]{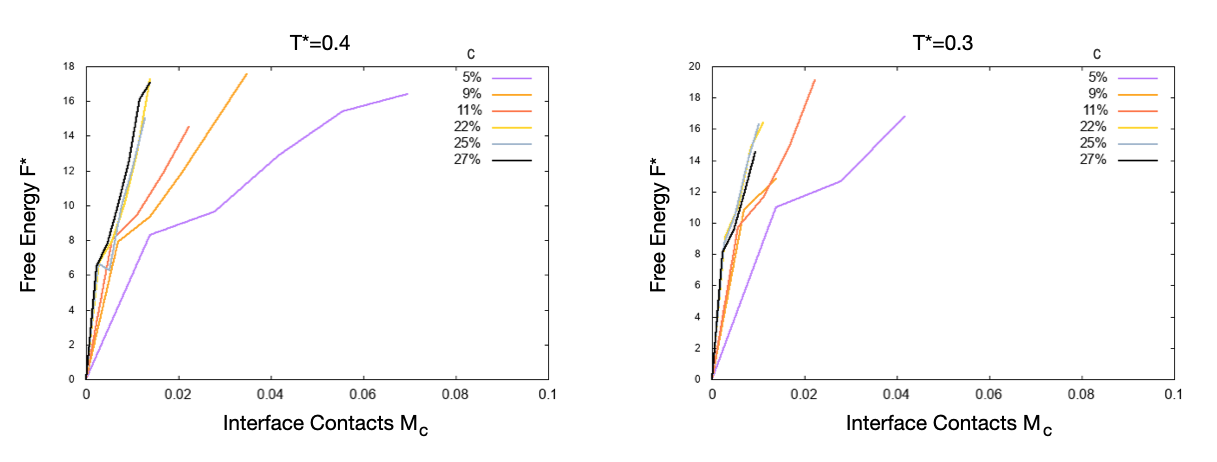}    
           \caption{Free energy profiles as a function of the normalized number, $M_c$, of~ contacts of the proteins with the hydrophobic interfaces.  
           Colors for the concentrations are as in Figure~\ref{fig:NcIc}. The~proteins are not adsorbed onto the interfaces when the minimum is at $M_c=0$.} 
       \label{fig:mem}
   \end{figure}

Indeed, our direct evaluation of the  free energy  as a function of the normalized number of
contacts of the proteins with the interface, $F^*(M_c)$,  
shows only minima at $M_c=0$ \mbox{(Figure~\ref{fig:mem})}.
 Hence, for~the specific sequence, $T$, and~$c$ we consider here, the~surface-adsorption of the proteins would have a free energy cost that is too large for the system. This~observation does not exclude that different sequences, with~larger numbers of hydrophobic residues 
 exposed to water in their native state, would adsorb at appropriate values of $T$ and $c$.
 
\begin{figure}[h!]
    \centering
   \includegraphics[scale=0.48]{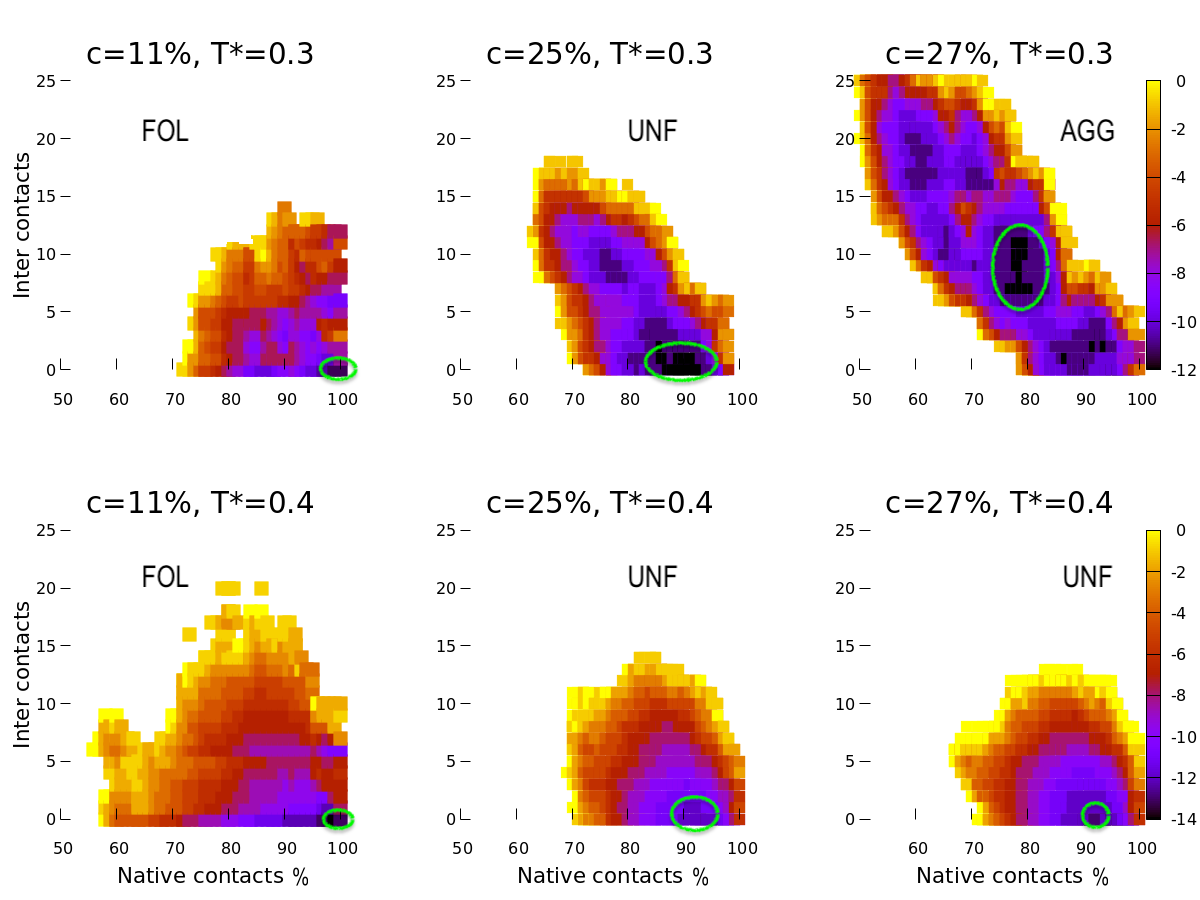}
     \caption{Color-coded free energy $F$ as a function of the native contacts $N_c$ (horizontal axis, expressed as a percentage) and the  average number of inter contacts $25N_p \times I_c$ (vertical axis), for~different concentrations and temperatures. Darker~colors correspond to deeper minima in free energy. We mark the absolute free-energy minima in green ellipses as guides for the eyes. We present data only for three concentrations. 
     At~ambient conditions (top panels) and $c=11\%$, the~proteins are mainly folded (FOL). 
     At~$c=25\%$, they are unfolded (UNF) on average.
     At $c=27\%$, they are unfolded and aggregated (AGG) in their majority.
     At~the warmer temperature (bottom panels),  they tend to avoid aggregation, even if they unfold. Statistically, where the volume concentration is low ($11\%$), the~proteins explore a larger number of  configurations than at higher density, given the high thermal energy and the larger available volume.}     
     \label{fig:landscape}
   \end{figure}

Considering that $M_c$ is not affecting  the free energy for our system in a significant way, 
we summarize our findings by the function $F^*(N_c,I_c)$, showing how the free energy depends on the two relevant parameters $N_c$ and $I_c$ (Figure~\ref{fig:landscape}). We find a clear correlation between the two at ambient conditions and high concentration, with~larger $I_c$ for smaller $N_c$. This result implies that at high concentration, $c\geq 25\%$, the~more the proteins unfold, the~more they~aggregate.

At lower $c$ and higher $T$, this correlation is weak.
In particular, $F^*(N_c,I_c)$  at higher $T$ (Figure~\ref{fig:landscape}, bottom)
emphasizes the propensity of the proteins to unfold only partially at higher $c$, without~aggregating.
Interestingly, at~the lower concentration in warm water,  there is a larger probability of~aggregation.

\subsection{Scale 1 and Scale~2}

Next, we explore 
 how unfolding, aggregation, and~surface-adsorption, depend on the competition between amino-acid interactions plus hydrophobic collapse, on~the one hand, and~entropy and energy-gain due to water--water HBs near hydrophobic interfaces, on~the other hand.
 To this goal, we reduce the latter,  by~changing the values of $\varepsilon^\Phi$, $J^\Phi$, and~$J_\sigma^\Phi$, as~indicated in Table~\ref{Tab1}, with~Scale 1 and Scale 2 sets of parameters compared to Scale~0.

\begin{figure}[h]
    \centering
    \includegraphics[scale=0.35]{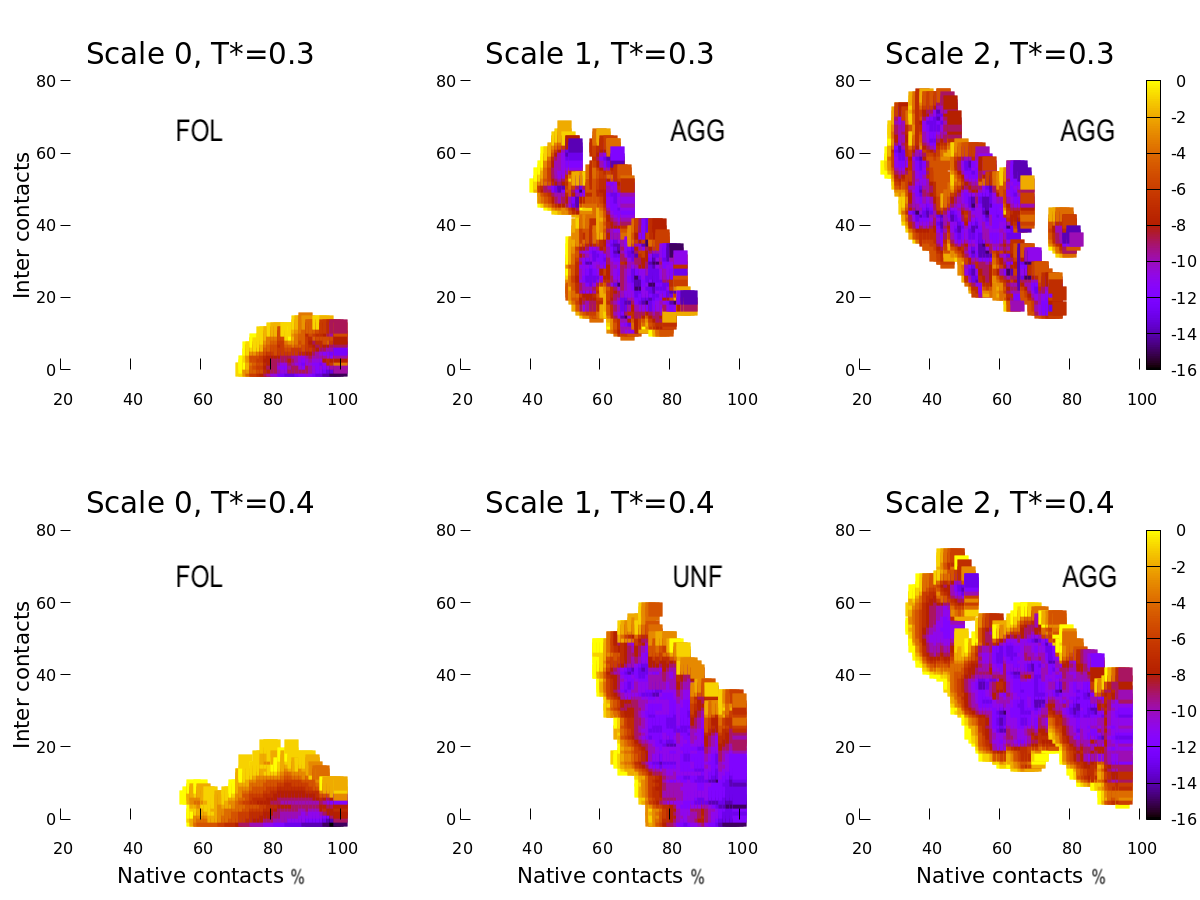}
     \caption{Comparing the free energy $F^*(N_c,I_c)$ for different choices of model parameters,   Scale 0, 1, 2  as  indicated in Table~\ref{Tab1}, at~fixed concentration $c=11\%$, and~ambient temperature (top row) or warmer temperature (bottom row). 
     Axes and colors are as in Figure~\ref{fig:landscape}. 
     Under these conditions, $F^*(N_c,I_c)$ for Scale 0 (left) has a minimum corresponding to folded proteins (FOL).
     For Scale 1 (center), at~ambient temperature (top), the~proteins are unfolded and aggregated (AGG), while at the higher temperature (bottom), they unfold (UNF) but have a small tendency to aggregate. 
          For Scale 2 (right), at~both temperatures, the~proteins aggregate (AGG) and are more unfolded at the lower temperature.}
     \label{fig:landscape_5prot}
   \end{figure}

We fix the concentration at $c=11\%$ and simulate the system at both temperatures considered for Scale 0, with~the parameters corresponding to Scale 1.
We observe that the proteins with Scale 1 parameters (Figure~\ref{fig:landscape_5prot}, central panels)
have a larger propensity to unfold and aggregate with respect to the model with the Scale 0 parameters \mbox{(Figure~\ref{fig:landscape_5prot}, left~panels)}. The~effect is even more evident with Scale 2 parameters (Figure~\ref{fig:landscape_5prot}, right panels).

In particular, at~warm temperature (Figure~\ref{fig:landscape_5prot}, bottom panels), the~weakening of the HB parameters in the $\Phi$-hydration shell increases the proteins' propensity to unfold (Scale~1) and aggregate (Scale 2). At~ambient temperature (Figure~\ref{fig:landscape_5prot}, top panels), both choices, Scale~1 and 2 lead to unfolded and aggregated~proteins. 

As a general trend, these calculations confirm that the aggregation propensity increases for the decreasing number of native contacts.  
     Although the heterogeneity of the plots reveals the difficulty to explore the accessible configurations for Scale 1 and 2 and to locate the absolute free-energy minimum, especially at lower $T$, the~overall trend of the results is clear. A~more thorough calculation of $F^*(N_c,I_c)$ is beyond the scope of the present~work.
     
\subsection{Effect of the Hydrophobic~Walls}
To elucidate the effect of the hydrophobic walls, we compare our results, at~ambient $T$,  with~those in bulk for the same protein~\cite{Bianco:2020aa}. We find that, at~least for this specific protein ($A_0$ in Ref.~\cite{Bianco:2020aa}), the~hydrophobic surfaces increase the concentration at which we observe unfolding, from $c_{\rm FOL \rightarrow UNF}\simeq 5\%$ to $11\%<c^{\rm S0}_{\rm FOL \rightarrow UNF}<22\%$, and~the aggregation threshold, from~ $c_{\rm UNF \rightarrow AGG}\simeq 20\%$ to $25\%<c^{\rm S0}_{\rm UNF \rightarrow AGG}<27\%$, where quantities with the superscript ${\rm S0}$ refer to the case presented here with the confining hydrophobic surfaces and Scale 0~parameters.

This result is apparently in contradiction with the findings in Ref.~\cite{Bianco-Navarro2019}. Bianco~et~al.  show that proteins  
tend to fold uninfluenced by the presence of other proteins provided that their single concentration is below their specific $c_{\rm FOL \rightarrow UNF}$~\cite{Bianco-Navarro2019}. This result could suggest that even the presence of an additional interface, such as the hydrophobic surfaces considered here, would not affect $c_{\rm FOL \rightarrow UNF}$. 

However, there are fundamental differences between the cases considered in Ref.~\cite{Bianco-Navarro2019} and the present work. While in Bianco~et~al.~\cite{Bianco-Navarro2019}, the additional interfaces (i) have a size comparable to the protein $A_0$, (ii) are heteropolymers made of different residues, and~(iii) are fluctuating in their positions and configurations, here they are (i) made of two infinite walls, (ii) homogenous in their hydrophobicity, and~(iii) fixed in~space. 

Hence, the~confining walls here exclude  a priori a number of protein configurations. On~the other hand, the~mixing with different proteins~\cite{Bianco-Navarro2019} alters the probability of some configurations for the protein $A_0$. However, it does not forbid any.
Therefore, we suggest that our confinement pushes the ${\rm FOL \rightarrow UNF}$ process toward protein concentrations that are higher than in bulk as a consequence of the limited ergodicity of the~system. 

While  our confining walls induce a 120\% increase in the  ${\rm FOL \rightarrow UNF}$ concentration  with respect to the bulk case~\cite{Bianco:2020aa}, our  ${\rm UNF \rightarrow AGG}$ concentration  is only 25\%  larger than the value in bulk~\cite{Bianco:2020aa}. This result is consistent with the fact that now the proteins unfold at a higher concentration. Hence, the~reduced free volume and the larger probability of protein--protein interaction partially compensate for the stabilization effect of the limited~ergodicity.

In all the cases we considered here, we find that the proteins make very few contacts with the walls and do not  adsorb onto them.  
We believe that this result is due mainly to the restrictions imposed by the 2D system, in~which the interface is reduced just to a line of~points. 

Furthermore,  the~sequence we consider here is mostly  (65\%) hydrophilic. 
Hence, the~proteins minimize their free energy when they are hydrated away from the~walls. 
 
We expect that proteins with larger hydrophobic patches would give rise to a more intense hydrophobic collapse,
associated with the larger bulk-water  entropy-gain and a stronger surface-adsorption. This investigation is~underway. 

More generally, it would be interesting to study how these results depend on the separation between the walls. 
For example,  the~system recovers the bulk case~\cite{Bianco-Navarro2019} at a large distance between the walls while 
at an intermediate distance, as~the one considered here, both ${\rm FOL \rightarrow UNF}$ and ${\rm UNF \rightarrow AGG}$ move to higher protein concentrations, with~no surface-adsorption.
The question is if these processes can still occur at smaller distances and how they would~change.

\subsection{Effect of the~Temperature}

We observe that temperature affects both unfolding and aggregation.
For the temperatures studied here, the~effect is small for the unfolding, 
without leading to a change in the ${\rm FOL \rightarrow UNF}$ threshold concentration.
However,  the~protein at low $c$ explores less unfolded states than at higher $T$, while they have   
a larger propensity to unfold  at higher $c$ and lower $T$.
In any case, at~lower $T$, the~system equilibrates more slowly and has a larger 
 statistical noise, especially at small $c$, because~of the smaller number of~proteins.

The effect is strong for the aggregation. While we find no aggregation at the warmer $T$ up to 
$c=27\%$, at~ambient $T$ it is $25\%<c^{\rm S0}_{\rm UNF \rightarrow AGG}<27\%$. 
We understand this result as a consequence of the larger propensity to unfold at lower $T$ and higher $c$, and~the thermal-energy decrease that favors the protein--protein interaction and~aggregation. 

Interestingly, at~the higher $T$, the~thermal energy hampers the aggregation of unfolded proteins, at~large $c$, more than that of folded proteins, at~small $c$. We interpret this finding as due to the larger steric hindrance of the loose ends of the unfolded proteins at high $T$. 

\subsection{Effect of the HB~Strength near a Hydrophobic Interface}

By going from Scale 0 to Scale 1 and 2, we reduce the HB strength near a hydrophobic interface. This~weakening would correspond in an experiment, e.g.,~to an increase of ions concentrations in the protein solution and a decrease of the hydrophobic effect, which contributes to the stability of the folded state against $T$ raises. 
Hence, at~fixed $T$ and $c$, the~FOL state is less stable when we go from Scale 0 to Scale 1 and from Scale 1 to Scale 2. This observation implies that  
$c^{\rm S0}_{\rm FOL \rightarrow UNF}\geq c^{\rm S1}_{\rm FOL \rightarrow UNF}\geq c^{\rm S2}_{\rm FOL \rightarrow UNF}$, consistent with our results about the propensity to unfold at fixed~concentration.

When we decrease the hydrophobic effect, the~relative importance of the protein--protein interaction increases, favoring the AGG state. Hence, we expect 
$c^{\rm S0}_{\rm UNF \rightarrow AGG}\geq c^{\rm S1}_{\rm UNF \rightarrow AGG}\geq c^{\rm S2}_{\rm UNF \rightarrow AGG}$, consistent with our~results.

As discussed for Scale 0, the~aggregation is more prominent at lower $T$  even for Scale 1 and 2.  In~particular, both Scale 1 and 2,  at~the chosen concentration $c=11\%$, lead directly to the AGG state. Hence, for~the set of parameters analyzed here, the~  
$c^{\rm S}_{\rm UNF \rightarrow AGG}(T^*=0.3)<c^{\rm S}_{\rm UNF \rightarrow AGG}(T^*=0.4)$, where the
superscript ${\rm S}$  refer to any of the scales for the parameters.
Further analysis will be necessary to verify all the above relations in detail for different sets of parameters and temperatures, including both confined and bulk~cases.

\section{Conclusions}

Following recent computational works, adopting the FS water model to  study how  folding/unfolding (FOL/UNF) compete with aggregation (AGG) when the protein concentration increases~\cite{Bianco:2020aa, 
Bianco-Navarro2019}, 
here we  consider the effect of nearby  hydrophobic walls at different temperatures, concentrations, and~ hydrophobic~strength. 

In all these cases, we find that the aggregation is ruled by the unfolding. The~more the proteins unfold, the~more they aggregate.
Increasing their concentration, the~proteins first unfold, ${\rm FOL \rightarrow UNF}$, 
and next aggregate, ${\rm UNF \rightarrow AGG}$,
with a range of concentrations for which the proteins unfold without aggregating, 
$c_{\rm FOL \rightarrow UNF}< c_{\rm UNF \rightarrow AGG}$~\cite{Bianco:2020aa}.

The presence of fixed hydrophobic walls increases both concentration thresholds at which  the processes  
${\rm FOL \rightarrow UNF}$ and 
${\rm UNF \rightarrow AGG}$ occurr, i.e.,~
$c_{\rm FOL \rightarrow UNF}<c^{\rm S}_{\rm FOL \rightarrow UNF}$, and~  
$c_{\rm UNF \rightarrow AGG}<c^{\rm S}_{\rm UNF \rightarrow AGG}$, with~a larger effect on FOL/UNF structural~change. 

We interpret these results as a consequence of the 
limitation of the accessible configurations that can be explored by the confined proteins (limited protein ergodicity). This effect is qualitatively  different from what Bianco~et~al. observed in simulations and experiments for bi-component protein solutions~\cite{Bianco-Navarro2019}, where protein--protein fluctuations alter the protein configurations probabilities but do not limit their~ergodicity. 

For the aggregation, the~decrease in free volume at large $c$ partially compensates the effect.
For the chosen geometry and specific  ({\it snake}) protein mainly hydrophilic, we do not observe surface-adsorption in the range of explored $T$ and $c$. Further analysis is underway for more hydrophobic~proteins.

Changes of $T$ also affect the limiting concentrations of FOL, UNF, and~AGG states. We find that at lower $T$ the {\it snake} proteins have a larger propensity to unfold and a much 
 stronger tendency to aggregate, as~a consequence of the decreased thermal energy. In~general, we find that 
$c^{\rm S}_{\rm UNF \rightarrow AGG}(T_{\rm low})<c^{\rm S}_{\rm UNF \rightarrow AGG}(T_{\rm high})$.

We find that changes in hydrophobic effect, as, e.g.,~due to an increase of ions in solution, 
also have a strong consequence on unfolding and aggregation. In~particular, we consider three HB strengths near hydrophobic interfaces, high (S0), intermediate (S1), and~small (S2), and~find results consistent with 
$c^{\rm S0}_{\rm FOL \rightarrow UNF}\geq c^{\rm S1}_{\rm FOL \rightarrow UNF}\geq c^{\rm S2}_{\rm FOL \rightarrow UNF}$,
and
$c^{\rm S0}_{\rm UNF \rightarrow AGG}\geq c^{\rm S1}_{\rm UNF \rightarrow AGG}\geq c^{\rm S2}_{\rm UNF \rightarrow AGG}$. Hence, the~decrease of the hydrophobic effect destabilizes the proteins against unfolding and aggregation at high concentrations, with~a stronger repercussion at lower $T$.

Our results are potentially useful for the understanding of the mechanisms that control  protein aggregation, a~process that is  associated with a growing
number of  neurodegenerative pathologies, including Alzheimer’s disease and Parkinson’s
disease~\cite{Brinieaaz3041,Mathieu:2020aa}.
They~represent the first step toward a multi-scale approach to study how to use nanostructured interfaces to regulate and, eventually, hamper  pathological protein aggregation~\cite{Vilanova:2017aa}. \par

\medskip
\textbf{Authors Contributions: } Each author has contributed as follows:
conceptualization, V.B. and G.F.; methodology, V.B. and G.F.; software, D.M. and V.B.; validation,  D.M.,  V.B. and G.F.; formal analysis, D.M.; investigation, D.M.,  V.B. and G.F.; resources, G.F.; data curation, D.M.; writing--original draft preparation, D.M.,  V.B. and G.F.; writing--review and editing, D.M.,  V.B. and G.F.; visualization, D.M.; supervision, V.B. and G.F.; project administration, G.F.; funding acquisition, G.F.
All authors have read and agreed to the published version of the~manuscript. \par

\textbf{Funding: } G.F. acknowledges the support of Spanish grant PGC2018-099277-B-C22 (MCIU/AEI /FEDER, UE), and the support by ICREA Foundation (ICREA Academia prize). \par

\end{document}